\documentclass[journal,twoside,web]{ieeecolor}
\let\labelindent\relax

\usepackage[utf8]{inputenc}
\usepackage[svgnames]{xcolor}
\usepackage{array,colortbl}
\usepackage{generic,graphicx}
\usepackage{textcomp}
\usepackage{amsmath} 
\usepackage{amssymb}  
\usepackage{amsfonts}
\usepackage{comment}
\usepackage{enumitem,here}
\usepackage{cases,xspace,enumitem}
\usepackage{stmaryrd}
\usepackage{lcsys}

\usepackage{tikz}
\usetikzlibrary{decorations.pathmorphing, arrows.meta}
\newcommand{\longsquigglyarrow}[1]{
\tikz \draw [->, line join=round, decorate, decoration={
    zigzag, segment length=4, amplitude=.9, post=lineto, post length=2pt
}] (0,0) -- (#1,0);
}

\usepackage{soul} 

\usepackage{version}
\includeversion{lcss}
\excludeversion{arxiv}

\graphicspath{{./epsfiles/},{./figures/}}

\usepackage{tikz}
\usetikzlibrary{positioning}

\usepackage[prependcaption,colorinlistoftodos]{todonotes}

\definecolor{gnred}{RGB}{255,91,89}

\definecolor{gnred1}{RGB}{71,0,0} 
\definecolor{gnred2}{RGB}{117,0,0} 
\definecolor{gnred3}{RGB}{164,0,0} 
\definecolor{gnred4}{RGB}{211,0,0} 
\definecolor{gnred5}{RGB}{255,0,0} 
\definecolor{gnred6}{RGB}{255,42,34} 
\definecolor{gnred7}{RGB}{255,91,89} 

\definecolor{gnblue1}{RGB}{0,36,71}   
\definecolor{gnblue2}{RGB}{0,60,118}  
\definecolor{gnblue3}{RGB}{0,85,164}
\definecolor{gnblue4}{RGB}{0,108,212}
\definecolor{gnblue4}{RGB}{0,108,212}
\definecolor{gnblue5}{RGB}{0,133,255}  
\definecolor{gnblue6}{RGB}{35,156,255} 
\definecolor{gnblue7}{RGB}{88,177,255} 

\definecolor{gnbrown1}{RGB}{71,27,0}  
\definecolor{gnbrown2}{RGB}{117,45,0} 
\definecolor{gnbrown3}{RGB}{164,62,0} 
\definecolor{gnbrown4}{RGB}{211,80,0} 
\definecolor{gnbrown5}{RGB}{255,97,0} 
\definecolor{gnbrown6}{RGB}{255,127,26} 
\definecolor{gnbrown7}{RGB}{255,155,86} 

\renewcommand{\theenumi}{(\roman{enumi})}

\newcommand\Item[1][]{%
  \ifx\relax#1\relax  \item \else \item[#1] \fi
  \abovedisplayskip=0pt\abovedisplayshortskip=0pt~\vspace*{-\baselineskip}}

\usepackage{hyperref}
\hypersetup{
    colorlinks=true,
    linkcolor=blue,
    filecolor=magenta,      
    urlcolor=cyan,
    }

\usepackage{amsthm}

\newtheoremstyle{ieeeconf}
  {0pt}   
  {0pt}   
  {\normalfont}  
  {\parindent}       
  {\itshape} 
  {:}         
  { } 
  {\thmname{#1} \thmnumber{#2}\thmnote{ (#3)}} 
\makeatletter
\renewenvironment{proof}[1][\proofname]{\par
  \pushQED{\qed}%
  \normalfont \topsep\z@
  \trivlist
  \item[\hskip2em
        \itshape
    #1\@addpunct{:}]\ignorespaces
}{%
  \popQED\endtrivlist\@endpefalse
}
\makeatletter

\theoremstyle{ieeeconf}

\newtheorem{property}{Property}

\theoremstyle{myplainroman}

\newcommand{\spec}{\operatorname{spec}}

\newcommand{\norm}[1]{\|#1\|}

\newcommand{\lognorm}[2]{\mu_{#2}(#1)}

\newcommand{\until}[1]{\{1,\dots, #1\}}

\newcommand{\subscr}[2]{#1_{\textup{#2}}}

\newcommand{\setdef}[2]{\{#1 \, | \, #2\}}

\newcommand{\map}[3]{#1 \colon #2 \rightarrow #3}

\newcommand{\osLip}{\operatorname{\mathsf{osL}}}
\newcommand{\osL}{\operatorname{\mathsf{osL}}}
\newcommand{\Lip}{\operatorname{\mathsf{Lip}}}

\newcommand{\snorm}[1]{{\left\vert\kern-0.25ex\left\vert\kern-0.25ex\left\vert #1 
		\right\vert\kern-0.25ex\right\vert\kern-0.25ex\right\vert}}

\newcommand{\xstar}{x^{*}}

\newcommand{\sign}{\operatorname{sign}}
\newcommand{\relu}{\operatorname{relu}}

\newcommand{\Iinfty}{I_{\infty}}

\DeclareSymbolFont{bbold}{U}{bbold}{m}{n}
\DeclareSymbolFontAlphabet{\mathbbold}{bbold}

\newcommand{\real}{\mathbb{R}}

\newcommand{\ds}{\displaystyle}
\newcommand{\e}{\mathrm{e}}

\newcommand{\mcE}{\mathcal{E}}

\newcommand{\mcK}{\mathcal{K}}

\newcommand{\mcKL}{\mathcal{KL}}

\newcommand{\msF}{\mathsf{F}}

\newcommand{\grad}{\nabla}

\newcommand{\fFR}{\subscr{\msF}{FR}}
\newcommand{\fG}{\subscr{\msF}{grad}}

\newcommand{\fGradCtrl}{\subscr{\msF}{GradCtrl}}

\newcommand{\msFLure}{\mathsf{F}_{\textup{Lur'e}}}

\newcommand{\jac}[1]{D\mkern-1mu{#1}}

\newcommand{\inprod}[2]{\left\langle\!\left\langle{#1},{#2}\right\rangle\!\right\rangle}

\newcommand{\change}[1]{#1}

\DeclareMathOperator*{\argmin}{arg\,min}

\newcommand{\myclearpage}{\clearpage}
\renewcommand{\myclearpage}{}

\begin{document}

\def\BibTeX{{\rm B\kern-.05em{\sc i\kern-.025em b}\kern-.08em
    T\kern-.1667em\lower.7ex\hbox{E}\kern-.125emX}}

\title{Contraction Theory for Optimization, Control, and Neural Networks}
\title{Perspectives on Contractivity \\ in Control, Optimization, and Learning}

\author{Alexander Davydov,~\IEEEmembership{Graduate Student~Member,~IEEE} \and
  Francesco Bullo,~\IEEEmembership{Fellow,~IEEE} \thanks{This work was in
    part supported by AFOSR project FA9550-21-1-0203 and NSF Graduate
    Research Fellowship under Grant 2139319.  The authors are with the
    Center for Control, Dynamical Systems, and Computation, UC Santa
    Barbara, Santa Barbara, CA 93106 USA. {\tt\small davydov@ucsb.edu,
      bullo@ucsb.edu}. All images are reprinted with permission
    from~\cite{FB:23-CTDS}.}}

\maketitle
\thispagestyle{empty}

\begin{abstract}
  Contraction theory is a mathematical framework for studying the
  convergence, robustness, and modularity properties of dynamical systems
  and algorithms. In this opinion paper, we provide five main opinions on
  the virtues of contraction theory. These opinions are (i) contraction
  theory is a unifying framework emerging from classical and modern works,
  (ii) contractivity is computationally-friendly, robust, and modular
  stability, (iii) numerous dynamical systems are contracting, (iv)
  contraction theory is relevant to modern applications, and (v)
  contraction theory can be vastly extended in numerous directions. We
  survey recent theoretical and applied research in each of these five
  directions.
\end{abstract}


\section*{Introduction}
\textit{Problem Description and Motivation:} 
\change{A discrete-time dynamical system is \emph{contracting} if its update map is a contraction in some metric. Analogously, a continuous-time system is contracting if its flow map is a contraction.}
Contraction theory for
dynamical systems is a set of concepts and tools for the study and design
of continuous and discrete-time dynamical systems. In this paper, we expose
a few comprehensive opinions on this field and, by doing so, we review the
basic theoretical foundations, the main computational and modularity
properties, three main example dynamical systems, some modern applications,
and extensions to local, weak, and Riemannian contraction.

\textbf{Opinion \#1: Contraction theory is a unifying framework emerging
  from classical and modern works.} Contraction theory originates from the
seminal work of Stefan Banach\change{\footnote{\change{The Banach fixed
    point theorem is also sometimes referred to as the
    Picard-Banach-Caccioppoli theorem in celebration of \'Emile Picard for
    his work on the ``method of successive approximations" in 1890 and the
    later independent work of Renato Caccioppoli~\cite{RC:1930}.}}}  in
1922~\cite{SB:1922}. Contraction mappings in continuous-time dynamical
systems can be traced back to the work of Lewis~\cite{DCL:49},
Demidovich~\cite{BPD:61}, and Krasovski\u {\i}~\cite{NNK:63}.  Logarithmic
norms and contraction for numerical integration of differential equations
was studied in the seminal works~\cite{GD:58, SML:58}. Later, logarithmic
norms were applied to control problems by Desoer and Vidyasagar
in~\cite{CAD-HH:72,CAD-MV:1975}. The term ``contraction analysis" was
coined in the seminal work by Lohmiller and Slotine where they studied
contraction with respect to Riemannian metrics~\cite{WL-JJES:98}.

It is now known that establishing contraction with respect to any norm has
equivalent differential tests (i.e., conditions on the Jacobian of the
vector field) and integral tests (i.e., conditions on the vector field
itself)~\cite{AD-SJ-FB:20o}. Before this unifying treatment, differential
and integral conditions for contractivity have been discovered and
rediscovered under different names. For example, focusing on integral
conditions the following eight notions are either identical or very closely
related:
\begin{enumerate}
\item one-sided Lipschitz maps in:
  \cite{GD:1976} and
  \cite{EH-SPN-GW:93}
  (Section~1.10, Exercise~6)      
\item {uniformly decreasing maps} in: \cite{LC-DG:76} 
\item no-name in: \cite{AFF:88} (Chapter~1, page~5)
\item {maps with negative nonlinear measure} in: \cite{HQ-JP-ZBX:01}
\item {dissipative Lipschitz maps} in: \cite{TC-PEK:05}
\item {maps with negative lub log Lipschitz constant} in: \cite{GS:06} 
\item {QUAD maps} in: \cite{WL-TC:06}
\item {incremental quadratically stable maps} in: \cite{LDA-MC:13}
\end{enumerate}


  
In other words, despite its deep historical roots, contraction theory is
still \change{awaiting uniformization and broader appreciation}. After decades of disparate work, a
comprehensive framework is now emerging that clarifies the relationship
among different strands of theoretical research.
  
Due to the non-uniformity in naming conventions, we argue that many
researchers are utilizing contraction-theoretic methods without placing
their work in the broader context of contraction theory. We believe that
contextualizing work in the framework of contraction theory can aid in
understanding many desirable properties of systems and in unifying their
mathematical treatment.

\textbf{Opinion \#2: Contractivity is computationally-friendly, robust, and
  modular stability.} Contracting dynamical systems exhibit highly ordered
transient and asymptotic behavior. Namely,
\begin{enumerate}
\item initial conditions are forgotten exponentially quickly and the
  distance between any two trajectories is monotonically decreasing;
\item a unique equilibrium is globally exponentially stable for a
  time-invariant vector field and two natural Lyapunov functions are
  automatically available;
\item when the vector field is time-varying and periodic, a unique periodic
  orbit exists and is globally exponentially stable; and
\item contracting dynamics enjoy highly robust behavior including (a)
  incremental input-to-state stability, (b) finite input-state gain, (c)
  contraction margin to unmodeled dynamics, and (d) input-to-state
  stability in the presence of delayed dynamics.
\end{enumerate}
Beyond their behavior, contracting dynamical systems admit systematic
procedures for the computation of their equilibria. Specifically, for a
given time-invariant contracting dynamical system, the forward Euler
integration of the dynamics with a specific step size guarantees that this
discrete-time iteration is also a contraction. Similar results can also be
proved for more sophisticated methods of numerical integration.

Finally, contraction theory is a modular framework. Given an
interconnection of contracting dynamical systems, conditions
exists~\cite{AD-SJ-FB:20o} under which the interconnection is contracting
and an explicit estimate of the contraction rate is available. In the case
of systems evolving on different time scales, contraction theory admits a
singular perturbation theorem that provides explicit estimates for
convergence rates and error estimates~\cite{LC-FB-EDA:23g}.

Due to the multitude of desirable consequences that contracting dynamics
enjoy, we believe that control theoreticians should search for
contractivity properties and design closed-loop systems that are
contracting.

\textbf{Opinion \#3: Numerous dynamical systems are contracting.} Fixed
point iterations are ubiquitous in algorithm design and can be seen as
discrete-time contracting dynamical systems. A classical example of a
discrete-time contracting dynamical system is the value iteration algorithm
from dynamic programming. Additionally, in convex optimization, many common
algorithms for finding minimizers of strongly convex functions including
gradient descent and projected gradient descent (for constrained
minimization) are precisely contracting dynamical systems. More generally,
in monotone operator theory, finding a zero of a strongly monotone operator
can be cast as finding a fixed point of a contractive map.

In online implementations, it is desirable to have continuous-time dynamics
that optimize a given objective. To this end, many of the contractive
iterative algorithms admit continuous-time analogs whereby the
continuous-time dynamics are contracting. Examples include gradient flow,
continuous-time primal-dual dynamics, and continuous-time proximal
gradient.

Beyond dynamics minimizing convex costs, there are many other
continuous-time contracting dynamical systems. A non-exhaustive list of
examples includes (i) stable LTI systems, (ii) neural networks (e.g.,
discrete and continuous-time, implicit and recurrent, etc) \change{with
  synaptic matrices satisfying certain norm conditions}, (iii)
exponentially incrementally input-to-state stable
dynamics~\cite{VA-BJ-LP:16}, (iv) Lur'e systems satisfying slope
constraints and certain LMI conditions~\cite{LDA-MC:13,MG-VA-ST-DA:23},
\change{stable monotone and positive systems satisfying certain
  conservation or invariance properties}~\cite{YK-BB-MC:20}, (vi) feedback
linearizable systems. More generally, it is known that any nonlinear system
with a locally exponentially stable equilibrium point is contracting with
respect to a suitable Riemannian metric~\cite{PG:15} inside the region of
attraction of the equilibrium point. Due to this multitude of examples of
contracting dynamics, we believe that the property of contraction is more
widespread than commonly thought and that researchers should actively seek
to establish contractivity properties.

\textbf{Opinion \#4: Contraction theory is relevant to modern
  applications.}  Beyond its prevalence in many common dynamical systems,
contraction theory is becoming increasingly relevant in modern control
applications as a design criterion. For example, in optimization and
control, there has been an influx of interest in controlling dynamical
plants by placing them in feedback with a controller which solves an
optimization problem in real-time. Examples of this paradigm are (i)
model-predictive control (MPC), (ii) online feedback optimization, and
(iii) methods based upon control barrier functions (CBFs). Rather than
assuming that the controller solves an optimization problem infinitely
fast, modern approaches study the co-evolution of system dynamics and
controller dynamics. As argued in~\cite{AK-ECB-AI-JL:23}, the key property
in establishing convergence of the co-evolution is incremental
input-to-state stability (which is implied by contractivity) and a
separation of time scale, which is readily analyzed via
contraction-theoretic tools. In online feedback optimization, tools based
upon contraction and singular perturbation were leveraged
in~\cite{LC-FB-EDA:23g}. Recently, contraction-theoretic ideas were even
applied to study the exponential stability of linear systems with
controllers designed via general parametric programs~\cite{AD-FB:24i}.

In addition to optimization and control, contraction-theoretic tools have
found manifold applications to machine learning and artificial neural
networks. For example, enforcing contractivity in neural ordinary
differential equations, implicit neural networks, and recurrent neural
networks has been demonstrated to improve the robustness of the models to
adversarial
perturbations~\cite{MR-RW-IRM:20,SJ-AD-AVP-FB:21f,MZ-LX-GFT:23,LK-ME-JJES:22}.
Beyond training robust neural networks, contractivity has also begun
pervading learning theory. As a first quintessential example, it was shown
in~\cite{SSD-XZ-BP-AS:19} that the error dynamics of overparametrized
neural networks are contracting with high probability. Additionally, it was
shown in~\cite{LK-PMW-JJES:23} that overparametrized neural networks
generalize well when they are trained with contractive optimizers (e.g.,
gradient descent).

While we have focused on applications to optimization, control and machine
learning, let us also mention the central role contractivity in dynamical
neuroscience~\cite{LK-ML-JJES-EKM:20,VC-FB-GR:22k,VC-AG-AD-GR-FB:23a},
robotics~\cite{SS-BL-AM-JJES-MP:23,HT-SJC:21b}, biochemical reaction
networks~\cite{GR-MDB-EDS:10a,MAAR-DA-EDS:23,AD-EDS:24}, cyberphysical
systems~\cite{GR-MDB:19} and
synchronization~\cite{WL-TC:06,PD-MdB-GR:11,ZA-EDS:14,MdB-DF-GR-FS:16}. Due
to the success that contraction theory has had in these application
domains, we believe that contraction theory can play equally as important a
role in many other modern applications.

\textbf{Opinion \#5: Contraction theory can be vastly extended in numerous
  directions.} Contractivity is a strong property
that implies many desirable consequences. Many practical applications
feature dynamical systems that are contracting in some generalized sense,
i.e., they still satisfy some of the desirable consequences that
contracting dynamics do. To capture these other classes of dynamical
systems, many extensions of contraction theory have been proposed. We
present a non-exhaustive list as follows: (i) contraction in Riemannian
metrics~\cite{WL-JJES:98,JWSP-FB:12za}, (ii) contraction of stochastic
systems~\cite{QCP-NT-JJES:09,ZA:22}, (iii) control contraction
metrics~\cite{IRM-JJES:17}, (iv) contraction on Finsler
manifolds~\cite{FF-RS:14}, (v) transverse contraction~\cite{IRM-JJES:14},
(vi) weakly contracting (or nonexpansive)
systems~\cite{SC:19,SJ-PCV-FB:19q}, (vii) semicontracting
systems~\cite{GDP-KDS-FB-MEV:21m}, (viii)
$k$-contraction~\cite{CW-IK-MM:22}, (ix)
$p$-dominance~\cite{FF-RS:19,YS-YK-NW:23}, (x) equilibrium
contraction~\cite{AD-SJ-FB:20o}, \change{and contraction for systems on time-scales~\cite{GR-FW:22}}.

Due to the vast number of extensions that contraction theory admits, we
believe that contraction theory and its applications will remain an active
area of research for years to come.

\textit{Paper Organization:} The paper is organized in five sections that
precisely elaborate our five main
opinions. Section~\ref{sec:contraction-basics} provides mathematical
preliminaries and establishes the unifying
framework. Section~\ref{sec:robust-comp-modular} highlights the
computationally-friendly, robust, and modular stability properties that
contracting dynamics enjoy. Section~\ref{sec:Numerous dynamical systems are
  contracting} provides examples of contracting dynamical
systems. Section~\ref{sec:modern-applications} highlights three modern
applications of contraction theory. Section~\ref{sec:extensions} provides a high-level overview of
some of the extensions to contraction theory. Finally, our last
Section~\ref{sec:open-problems} contains conjectures and open problems for
future research.

\myclearpage

\section{Contraction theory is a unifying framework emerging from classical and modern works}
\label{sec:contraction-basics}

In this section we review some basic definitions, properties and examples
of contracting dynamical systems. We focus on systems defined on
finite-dimensional vector spaces with norms. The section leads to a table
of contractivity conditions, namely Table~\ref{table:osLip}, that unifies
numerous previous results. In what follows we silently assume sufficient
smoothness of all objects.

\smallskip\subsubsection*{Norms and induced norms}
\begin{figure}[ht]\centering
  \includegraphics[width=.75\linewidth]{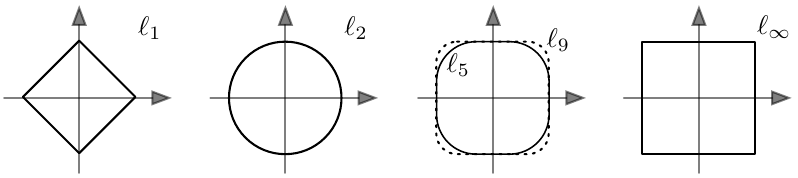}
  \caption{The unit disks $\setdef{v\in\real^2}{\norm{v}{}=1}$ in the
    $\ell_1$, $\ell_2$, $\ell_p$ for $p\in\{5,9\}$, and $\ell_\infty$
    norms. } \label{fig:Holder-unitcircles}
\end{figure}

\begin{figure}[h]
  \includegraphics[width=.44\linewidth]{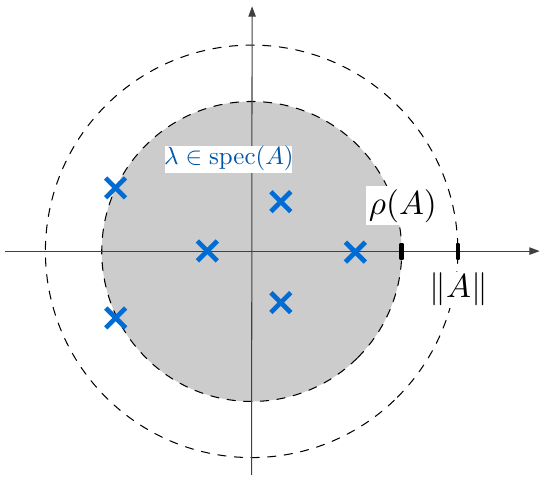}
  \hfill \includegraphics[width=.44\linewidth]{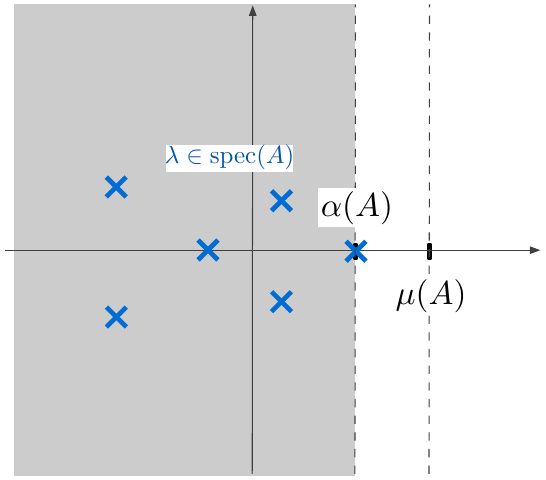}
  \caption{Given $A\in\real^{n\times n}$ with spectrum $\spec(A)$, the
    induced norm $\norm{A}$ is an upper bound on the spectral radius
    $\rho(A)$ (left panel) and the induced log norm $\lognorm{A}{}$ is an
    upper bound on the spectral abscissa $\alpha(A)$ (right
    panel).}\label{fig:complexplane}
\end{figure}

Classic norms on $\real^n$ are the $\ell_p$ norms for $p\in[1,\infty]$,
given by $\norm{v}_{1} = \sum_{i=1}^n |v_i|$, $\norm{v}_{2} =
\sqrt{\sum_{i=1}^n v_i^2}$, $\norm{v}_{\infty} = \max_{i\in\until{n}}
|v_i|$, and $\norm{v}_{p}=\big(\sum\nolimits_{i=1}^n |v_i|^p\big)^{1/p}$
for every other value of $1<p<\infty$; see
Figure~\ref{fig:Holder-unitcircles}.
Given a positive definite $P$ and a positive vector $\eta$, we also define
$\norm{x}_{2,P}=(x^\top P x)^{1/2}$ and
$\norm{x}_{\infty,\eta}=\max_i|x_i|/\eta_i$.  Given $A\in \real^{n\times
  n}$, a norm $\norm{\cdot}$ on $\real^n$ induces a \emph{matrix norm} and
a \emph{matrix log norm} defined by
\begin{align*}
  \norm{A} := \max_{\norm{x}=1}\norm{Ax} \quad\text{and}\quad
  \lognorm{A}{} := \lim_{h\to 0^+}\frac{\|I_n + hA\|-1}{h}.
\end{align*}
Closed form expressions are known for matrix norms and matrix log norms for
all norms defined above, e.g., see~\cite[Chapter~2]{FB:23-CTDS}.  Matrix
norms and matrix log norms enjoy a wide range of properties, including
subadditivity, positive scaling, convexity, as well as certain monotonicity
properties. Figure~\ref{fig:complexplane} illustrates how they are related
to the eigenvalues of $A$.

\smallskip\subsubsection*{Discrete-time contractivity via Lipschitz constants}
We consider discrete-time dynamical systems and their contractivity via
Lipschitz constants. We consider
\begin{equation*}
  \text{ $x_{k+1}=\msF(x_k)$ on $\real^n$ with norm $\norm{\cdot}{}$ and
    induced norm $\norm{\cdot}{}$ }
\end{equation*}
The \emph{Lipschitz constant} of $\msF$ is equivalently defined by
\begin{align*}
  \Lip(\msF) &= \inf \setdef{\ell>0\,}{ \,\norm{\msF(x)-\msF(y)}{} \leq
    \ell \norm{x-y}{} \text{ for all } x,y } \\ &= \sup\nolimits_x \norm{
    \jac{\msF}(x)}{}
\end{align*}
For a scalar map $\map{f}{\real}{\real}$, we have $\Lip(f)=\sup_x |\change{f'(x)}|$.
Given a scalar $\ell$ and an affine map $\msF_A(x)=Ax+a$, for
$A\in\real^{n\times{n}}$ and $a\in\real^n$, we note the equivalences:
\begin{align*}
  \Lip_{2,P}(\msF_A)=\norm{A}_{2,P}\leq\ell\qquad&\iff\qquad A^\top P A \preceq \ell^2 P     \\
  \Lip_{\infty,\eta}(\msF_A) = \norm{A}_{\infty,\eta} \leq \ell \qquad&\iff\qquad \change{|A|\eta \leq \ell \eta},
\end{align*}
\change{where $|A|$ denotes the entrywise absolute value of $A$.}
Loosely-speaking, the computation of the Lipschitz constant of an affine
map is a semidefinite feasibility program in the weighted
$\ell_2$ case and a linear program in the weighted $\ell_\infty$ case. With
these concepts at hand, the \emph{Banach contraction theorem for
discrete-time dynamics} states that, if $\rho:=\Lip(\msF)<1$, then
\begin{enumerate}
\item the map $\msF$ is strongly contracting, i.e., the distance between
  any two trajectories decreases exponentially fast (with contraction
  factor $\rho$), and
\item $\msF$ has a globally exponentially stable equilibrium~$x^*$.
\end{enumerate}
\begin{figure}[h]\centering
  \includegraphics[width=.75\linewidth]{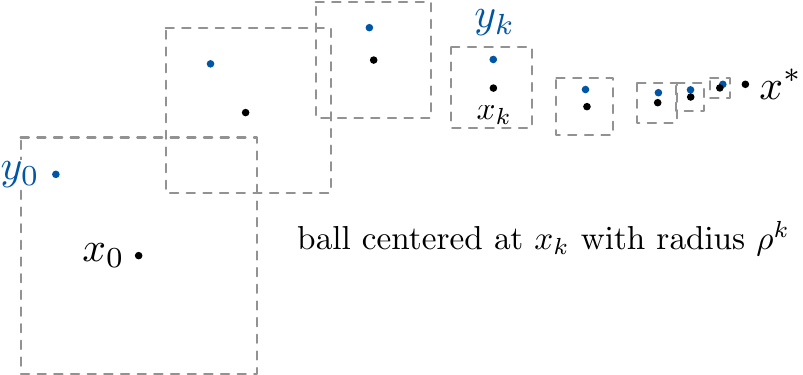}
  \caption{The Banach contraction theorem for discrete-time dynamics: Given
    initial conditions $x_0$, $y_0$ with $\norm{x_0-y_0}\leq1$, $y_k$
    remains inside the ball centered at $x_k$ with radius $\rho^k$ and both
    trajectories converge exponentially fast to a limiting point $x^*$.}
\end{figure}

\begin{table*}[t]
	\centering\small
	\resizebox{1\textwidth}{!}{\begin{tabular}{
				p{0.19\linewidth}
				p{0.34\linewidth}
				p{0.4\linewidth}
			}
			Log norm & differential & \textcolor{blue}{one-sided Lipschitz} \\
			condition &  condition &  condition \\
			\hline
			\rowcolor{LightGray}    & &  \\[-1ex]
			\rowcolor{LightGray}
			$ \ds \mu_{2,P}(\jac{\msF}(x))\leq b$ &
			$\ds P \jac{\msF}(x) + \jac{\msF}(x)^\top  P \preceq 2 b P $
			& $\ds (x-y)^\top  P \big( \msF(x) - \msF(y) \big) \leq b \norm{x-y}_{P^{1/2}}^2$
			\\[2ex]
			\rowcolor{White} && \\[-1ex]
			\rowcolor{White}
			$\ds \mu_{1}(\jac{\msF}(x))\leq b$
			&  $\ds \sign(v)^{\top} \jac{\msF}(x) v\le b \norm{v}_{1}$
			&  $\ds \sign(x-y)^{\top} (\msF(x) - \msF(y))\le b \norm{x-y}_{1}$
			\\[2ex]
			\rowcolor{LightGray} && \\[-1ex]
			\rowcolor{LightGray}
			$\ds \mu_{\infty}(\jac{\msF}(x))\leq b$
			&  $\ds     \max_{i\in \Iinfty(v)}\! v_i \left(\jac{\msF}(x) v\right)_i
			\le b \norm{v}_{\infty}^2$ 
			&
			$\ds\max_{i\in\Iinfty(x-y)} \! (x_i-y_i) (\msF_i(x)-\msF_i(y)) \leq b \norm{x-y}_{\infty}^2$
			\\[2ex]\hline
	\end{tabular}}
	\caption{Table of equivalent contractivity conditions for a vector field
		$\map{\msF}{\real^n}{\real^n}$ from~\cite{AD-SJ-FB:20o}.
		Each row contains three equivalent statements for the
                weighted $\ell_2$, $\ell_1$, and $\ell_\infty$ norm,
                respectively. Each statement is to be understood for all
                $x,y\in\real^n$ and all $v\in\real^n$.
		Equation~\eqref{def:osLip} corresponds to the first column. The second
		column contains generalized version of the classic Demidovich condition
		for the existence of a common Lyapunov function.  The third column
		contains ``integral'' conditions that do not require differentiability of $\msF$.
		The function $\map{\sign}{\real^n}{\{-1,0,1\}^n}$ is the entrywise
		$\sign$ function. We adopt the shorthand $\Iinfty(v) =
		\setdef{i\in\until{n}}{|v_i|=\norm{v}_{\infty}}$.
	} \label{table:osLip}
\end{table*}

\smallskip\subsubsection*{Continuous-time contractivity via one-sided Lipschitz constants}
Next we present a perfectly parallel treatment of the continuous-time case.
We consider continuous-time dynamical systems and their contractivity via
one-sided Lipschitz constants. We consider
\begin{equation*}
  \text{
    $\dot{x}=\msF(x)$ on $\real^n$ with norm
    $\norm{\cdot}{}$ and induced log norm $\lognorm{\cdot}{}$
  }
\end{equation*}
The \emph{one-sided Lipschitz constant} of $\msF$ is defined by
\begin{align}
  \label{def:osLip}
  \osLip(\msF) &= \sup\nolimits_x \lognorm{ \jac{\msF}(x)}{}
\end{align}
and we refer to Table~\ref{table:osLip} for more general and equivalent
definitions.  For a scalar map $\map{f}{\real}{\real}$, we have
$\osLip(f)=\sup_x f'(x)$.  Given a scalar $\ell$ and an affine map
$\msF_A(x)=Ax+a$, for $A\in\real^{n\times{n}}$ and $a\in\real^n$, we note
the equivalences:
\begin{align*}
  \osLip_{2,P}(\msF_A)=\lognorm{A}{2,P}\leq \ell \enspace&\iff\enspace
  A^\top P +AP \preceq 2 \ell P       \\
  \osLip_{\infty,\eta}(\msF_A) = \lognorm{A}{\infty,\eta} \leq \ell \enspace&\iff\enspace
  a_{ii} + \sum_{j\neq i}|a_{ij}|\eta_i/\eta_j \leq \ell
\end{align*}
Again, the computation of the one-sided Lipschitz constant of an affine map
is equivalent to a semidefinite feasibility program in the weighted
$\ell_2$ case and a linear program in the weighted $\ell_\infty$ case. With
these concepts at hand, the \emph{\change{contraction} theorem for
continuous-time dynamics} states that, if $-c:=\osLip(\msF)<0$, then
\begin{enumerate}
\item the vector field $\msF$ is strongly infinitesimally contracting,
  i.e., the distance between any two trajectories decreases exponentially
  fast (with contraction rate $c$), and
\item $\msF$ has a globally exponentially stable equilibrium~$x^*$.
\end{enumerate}

\begin{figure}[h]\centering
  \includegraphics[width=.75\linewidth]{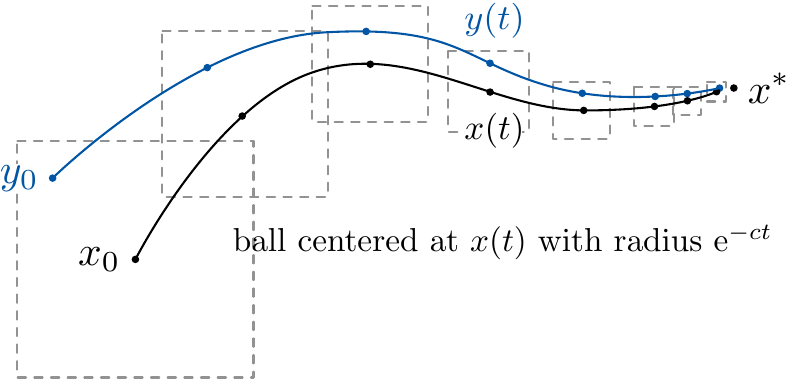}
  \caption{\change{Contraction} theorem for continuous-time dynamics:
    Given initial conditions $x_0$, $y_0$ with $\norm{x_0-y_0}\leq1$,
    $y(t)$ remains inside the ball centered at $x(t)$ with radius
    $\e^{-ct}$ and both trajectories converge exponentially fast to a
    limiting point $x^*$.}
\end{figure}

\change{Mathematically, for the continuous-time dynamics $\dot{x} =
  \msF(x)$ with $\osL(\msF) = -c < 0$, from any two initial conditions,
  $x_0,y_0$, the corresponding solutions $x(\cdot),y(\cdot)$ satisfy
\begin{equation}\label{eq:incremental-stability}
	\|x(t)-y(t)\| \leq e^{-ct}\|x_0-y_0\|.
\end{equation}
This bound is a special type of incremental stability,~\cite{DA:02}, namely
\emph{incremental exponential stability without overshoot}. Additionally,
the incremental Lyapunov function $V(x,y) = \|x-y\|$ can be used to
establish the bound~\eqref{eq:incremental-stability}.
}

\smallskip\subsubsection*{Table of contractivity conditions}
As promised at the beginning of this section, we are finally ready to
provide a unifying table of contractivity conditions emerging from
classical and modern works. Table~\ref{table:osLip}, taken
from~\cite{AD-SJ-FB:20o}, see also~\cite{ZA-EDS:14b}, provides three
equivalent continuous-time contractivity conditions (namely, log norm,
\change{differential, and one-sided Lipschitz} conditions) for the $\ell_2$, $\ell_1$ and
$\ell_\infty$ norms.  The table summarizes the following classical and
modern works:
\begin{itemize}
\item log norm condition (each row, first entry): classic
  work~\cite{CAD-HH:72};
\item $\ell_2$ differential condition (first row, second entry): classic
  work~\cite{BPD:61,NNK:63} and its Riemannian
  generalization~\cite{WL-JJES:98};
\item $\ell_2$ integral condition (first row, third entry): see the
  discussion in the Introduction;
\item $\ell_1$ integral condition (second row, third entry): due to work on
  neural networks and traffic networks \cite{HQ-JP-ZBX:01,GC-EL-KS:15};
\item $\ell_\infty$ conditions (third row): second and third entry are only
  very recently discovered~\cite{AD-SJ-FB:20o}.
\end{itemize}

\change{We conclude this section by remarking that while Table~\ref{table:osLip} only studies contraction conditions for continuous-time dynamics, another comparable table can be created for contraction conditions for discrete-time dynamics based upon the induced norm and Lipschitz conditions.}



\section{Contractivity is robust, computationally-friendly, and modular stability}\label{sec:robust-comp-modular}
We now present some selected properties enjoyed by strongly contracting
dynamical systems. We mostly focus on the continuous-time case\change{; we comment on the discrete-time case at the end of this section}. To illustrate our Opinion
\#2, we present robustness, computational, and modularity properties.

We start with a well-known robustness property~\cite{WL-JJES:98} with
respect to input disturbances and a recent
extension~\cite{AD-VC-AG-GR-FB:23f} to the tracking of equilibrium
trajectories.

\begin{figure}[h]\centering
  \includegraphics[width=.95\linewidth]{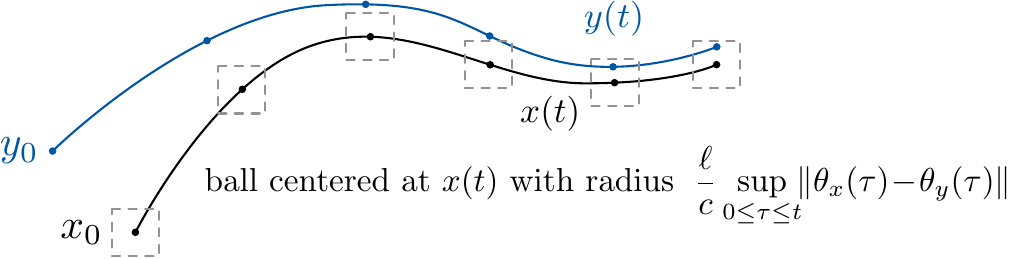}
  \caption{The incremental ISS property: each two trajectories $x(t)$ and
    $y(t)$ determined by inputs $\theta_x(t)$ and $\theta_y(t)$
    respectively, asymptotically reach a distance upper bounded by
    $\frac{\ell}{c}\sup_{\tau}\norm{\theta_x(\tau)-\theta_y(\tau)}{}$.}\label{fig:contractivity-iISS}
  
  \includegraphics[width=.95\linewidth]{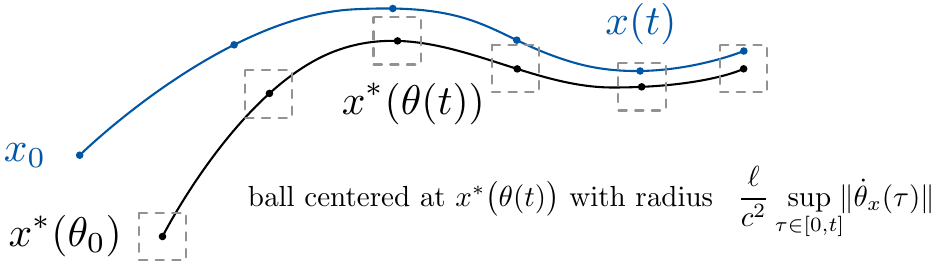}
  \caption{The equilibrium tracking property: each trajectory $x(t)$
    asymptotically reaches a distance from the equilibrium trajectory
    $x^*\bigl(\theta(t)\bigr)$ upper bounded by
    $\frac{\ell}{c^2}\sup_{\tau}\norm{\dot{\theta}(\tau)}{}$.}\label{fig:contractivity-eqTracking}
\end{figure}

\begin{property}[Incremental input-to-state stability (iISS) and equilibrium tracking]
  \label{prop:iISS+eqT}
  Consider a dynamical system subject to an input $\dot{x} =
  \msF(x,\theta(t))$. Assume
  \begin{itemize}
  \item\label{ass:iss:contract} (contractivity with respect to $x$:) $\ds
    \osLip_x(\msF) \leq -c <0$, uniformly in $\theta$, and
  \item\label{ass:iss:lipsch} (Lipschitz with respect to $\theta$:) $\ds
    \Lip_\theta(\msF) \leq \ell$, uniformly in $x$.
  \end{itemize}  
   Then the system enjoys the \emph{incrementally ISS property} (see
   Figure~\ref{fig:contractivity-iISS}), namely for any two trajectories
   $x(t)$ and $y(t)$,
  \begin{multline}\label{eq:iiss}
      \|x(t) - y(t)\| \leq \e^{-ct}\|x_0 - y_0\| \\ +
      \frac{\ell}{c}(1-\e^{-ct})
      \sup_{\tau \in [0,t]}\norm{\theta_x(\tau)-\theta_y(\tau)}{}. 
  \end{multline}
  Additionally, at fixed $\theta$, let $x^*(\theta)$ be the unique of
  equilibrium of $\dot{x}=\msF(x,\theta)$.  Then the system enjoys the
  \emph{equilibrium tracking property} (see
  Figure~\ref{fig:contractivity-eqTracking}), namely
  \begin{multline*}
      \|x(t){-}\xstar(\theta(t))\| \leq \e^{-ct}\|x_0{-}\xstar(\theta_0)\|
      \\ + \frac{\ell}{c^2} (1-\e^{-ct})
      \sup_{\tau\in[0,t]}\|\dot\theta(\tau)\|
  \end{multline*}
\end{property}

\change{The incremental ISS bound~\eqref{eq:iiss} is a special case of the more general incremental ISS bound~\cite[Definition~4.1]{DA:02}
\begin{equation*}
	\|x(t)-y(t)\| \leq \beta(\|x_0-y_0\|,t) + \gamma(\sup\nolimits_{\tau \in [0,t]} \|\theta_x(\tau)-\theta_y(\tau)\|),
\end{equation*}
where $\beta$ is a $\mcKL$ function and $\gamma$ is a $\mcK_\infty$
function. In other words, contraction affords an exponential-form
incremental ISS whereby transient terms decay exponentially.  }

Next, we present a computational result explaining a close relationship
between continuous and discrete-time contracting systems.
\begin{property}[Euler discretization]
  Given arbitrary norm $\norm{\cdot}{}$ and Lipschitz
  $\map{\msF}{\real^n}{\real^n}$, the following statements are equivalent:
  \begin{enumerate}
  \item\label{p1:osLF} $\dot{x}=\msF(x)$ is strongly infinitesimally
    contracting, and
  \item\label{p2:LipFalpha} there exists $\alpha>0$ such that
    $x_{k+1}=x_k+\alpha\msF(x_k)$ is strongly contracting.
  \end{enumerate}
\end{property}

Guidelines for the selection of an optimal step-size are available: the
optimal choice depends upon the norm. We refer to~\cite{FB-PCV-AD-SJ:21e} for guidance on step-size selection. Finally, we present a result about
the modularity properties of contracting systems.
    
\begin{property}[Network contractivity]\label{prop:network}
  Consider $n$ interconnected subsystems described by
  \begin{align*}
    \dot{x}_i = \msF_i (x_i, x_{-i}), \qquad \text{for } i\in\until{n}
  \end{align*}
  where $x_i\in\real^{N_i}$ is the state of the $i$th subsystem and
  $x_{-i}\in\real^{N-N_i}$ are the states of all other subsystems. Assume
  that
  \begin{itemize}
  \item (contractivity with respect to $x_i$:) $\ds \osLip_{x_i}(\msF_i) \leq
    -c_i <0$, uniformly in $x_{-i}$

  \item (Lipschitz with respect to $x_j$, $j\neq{i}$:) $\ds
    \Lip_{x_j}(\msF_i) \leq \ell_{ij}$, uniformly in $x_{-j}$, and
  
\item the gain matrix $\Gamma=\begin{bmatrix} -c_1 & \dots & \ell_{1n}\\ \vdots &
  & \vdots \\ \ell_{n1} & \dots & - c_n
  \end{bmatrix}$ is {Hurwitz}.
 \end{itemize}
Then the interconnected system is strongly infinitesimally contracting with
rate equal to $|\alpha(\Gamma)|$.
\end{property}
Note that the Hurwitz assumption on the Metzler matrix $\Gamma$ is
equivalent to a small gain theorem~\cite{XD-SJ-FB:19f} and that there are
other interconnection theorem based upon times scale
separation~\cite{DDV-JJES:13,LC-FB-EDA:23g}.

\change{It is useful to note that Properties~\ref{prop:iISS+eqT}
  and~\ref{prop:network} admit natural analogous results for discrete-time
  contracting dynamics. Namely, discrete-time contracting dynamics with
  input are naturally incrementally ISS and a network of discrete-time
  dynamics are contracting if a corresponding gain matrix is \emph{Schur
  stable} rather than Hurwitz.}

\section{Numerous dynamical systems are contracting}
\label{sec:Numerous dynamical systems are contracting}
The simplest example or a discrete-time (resp.,~continuous-time)
contracting dynamics is the case of a linear system with a Schur
(resp,~Hurwitz) matrix. This is due to the existence of efficient norms,
see~\cite[Chapter~2]{FB:23-CTDS}.  This observation can be easily extended
to feedback linearizable systems with stabilizing controllers.  We now
present three important examples of continuous-time contracting dynamics
from convex optimization, neural networks, and nonlinear control.

\smallskip\subsubsection*{Gradient descent in optimization}
Given a function $\map{f}{\real^n}{\real}$, the following are equivalent
statements:
\begin{enumerate}
\item\label{SCfact:def} $f$ is \emph{strongly convex} with parameter $\nu$
  (and global minimum $x^*\in\real^n$), and
\item\label{fact:stronglyconvex-contracting} the \emph{gradient descent}
  vector field $\fG := -\grad{f}$ is strongly infinitesimally contracting
  with respect to $\norm{\cdot}_2$ with rate $\mu$ (and equilibrium $x^*$).
\end{enumerate}
This result is also known as Kachurovskii's Theorem~\cite{RIK:60}.  Many
other related dynamical systems are contracting~\cite{AG-AD-FB:23o} under
strong convexity assumptions, including primal-dual gradient descent,
incidence- and Laplacian-based distributed gradient descent, proximal
gradient descent as well as saddle dynamics, pseudo-gradient descent, and
best response play from game theory.

\change{The contractivity of ${-}\nabla f$ when $f$ is strongly convex is a
  property of a continuous-time dynamical system. To the best of our
  knowledge, there is no natural discrete-time contraction property to
  assess whether a given function is strongly convex.}
    
\smallskip\subsubsection*{Firing rate models in recurrent neural networks}
\newcommand{\slope}[2]{\mathrm{slope}[#1,#2]}
\newcommand{\dmin}{\subscr{d}{1}}
\newcommand{\dmax}{\subscr{d}{2}}

\begin{figure}[h]\centering
  \includegraphics[width=0.46\linewidth,height=.2\textheight,keepaspectratio]{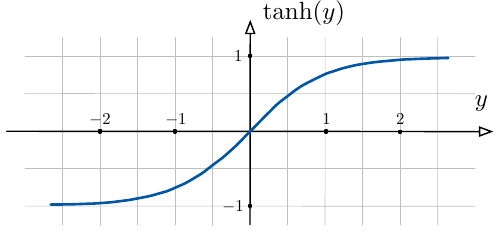}\quad
  \includegraphics[width=0.46\linewidth,height=.2\textheight,keepaspectratio]{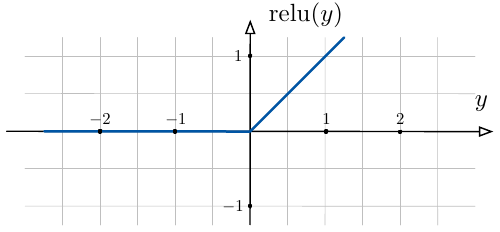}
  \caption{Hyperbolic tangent and relu functions satisfy the constraint
    $\dmin=0\leq\phi_i'(y)\leq 1=\dmax$ almost everywhere.} \label{fig:activations}
\end{figure}

Next, we consider the \emph{firing rate model} for recurrent neural networks
\begin{equation}\label{eq:firingrate}
  \dot{x} = -Cx + \Phi(Ax + u) =: \fFR(x).
\end{equation}
where the dissipation matrix $C \in \real^{n \times n}$ is diagonal and
positive semi-definite, the synaptic matrix $A\in \real^{n \times n}$ is
invertible, $u \in \real^n$ is a constant bias, and
$\map{\Phi}{\real^n}{\real^n}$ is an activation function for the form
$\Phi(x) = (\phi_1(x_1), \dots, \phi_n(x_n))$, where each
$\map{\phi_i}{\real}{\real}$ satisfies
\begin{equation}\label{eq:slope-restricted}
  \begin{aligned}
    \dmin & := \inf_{x,y\in \real, x\neq y}\frac{\phi_i(x) - \phi_i(y)}{x - y}, \\
    \dmax & := \sup_{x,y\in \real, x\neq y}\frac{\phi_i(x) - \phi_i(y)}{x - y}.
  \end{aligned}
\end{equation}
for appropriate finite $\dmin\leq\dmax$, see the examples in
Figure~\ref{fig:activations}.  Then
\begin{equation*}
  \osL_{\infty}(\fFR)=\max\{ \mu_{\infty}(-C+\dmin A), \mu_{\infty}
  (-C+\dmax A)\},
\end{equation*}
In other words, when $\dmin=0$, $\dmax=1$, $C=I_n$, and
$\relu\bigl(\mu_{\infty}(A)\bigr)<1$, we know
\begin{equation*}
  \osL_{\infty}(\fFR)=-1+\relu\bigl(\mu_{\infty}(A)\bigr)
\end{equation*}
so that the firing rate model~\eqref{eq:firingrate} is strongly infinitesimally
contracting with respect to $\|\cdot\|_{\infty}$ with rate
$1-\relu\bigl(\mu_{\infty}(A)\bigr)$.

This result follows from maximizing a convex function on a polytope. We
refer to~\cite{AD-AVP-FB:22q} for a comprehensive sharp treatment of
contractivity of continuous-time recurrent neural networks with respect to
optimally-weighted non-Euclidean norms.

\smallskip\subsubsection*{Lur'e models in nonlinear control}
Given matrices $A\in\real^{n\times{n}}$, $B\in\real^{n\times{m}}$,
$C\in\real^{m\times{n}}$ and a map $\map{\varphi}{\real^m}{\real^m}$,
consider the \emph{Lur'e system}
\begin{equation}\label{eq:Lure}
  \dot{x} = Ax + B\varphi(Cx) :=\msFLure(x).
\end{equation}
We consider maps $\varphi$ for which $\rho>0$ exists such that
\begin{equation}\label{eq:monotonicity}
  (\varphi(y_1)-\varphi(y_2))^\top(y_1-y_2) \geq \rho\|\varphi(y_1)-\varphi(y_2)\|_2^2
\end{equation}
for all $y_1,y_2 \in \real^m$ (such maps are referred to as
\emph{cocoercive}).  Given a positive definite $P \in \real^{n \times n}$
and $\eta>0$, the following statements are equivalent:
\begin{enumerate}
\item the Lur'e system~\eqref{eq:Lure} is strongly infinitesimally
  contracting with respect to $\|\cdot\|_{2,P}$ with rate $\eta$ for any
  $\varphi$ satisfying~\eqref{eq:monotonicity},
\item there exists $\lambda \geq 0$ such that
  \begin{equation} \label{eq:LMI-Lure}
    \begin{bmatrix}
      A^\top P + PA + 2\eta P & PB + \lambda C^\top \\ B^\top P + \lambda C & -2\lambda \rho I_m
    \end{bmatrix} \preceq 0.
  \end{equation}
\end{enumerate}
In other words, the LMI~\eqref{eq:LMI-Lure} is necessary and sufficient for
\begin{equation*}
  \sup_{\varphi \text{ satisfying } \eqref{eq:monotonicity}} \osL(\msFLure) = -\eta.
\end{equation*}
This result follows from a careful application of the S-lemma.





\myclearpage

\section{Contraction theory is relevant to modern applications}\label{sec:modern-applications}
We here briefly review three application areas.

\smallskip\subsubsection*{Optimization-based control}
We argue that contraction theory is relevant in a wide range of
optimization and optimization-based control problems, e.g., including
parametric optimization, model predictive control~\cite{AK-ECB-AI-JL:23},
control barrier functions, and online feedback optimization, as we show
below. We focus on continuous-time systems\change{, but comment on discrete-time ones when there are key distinctions.}

Since many convex optimization problems can be solved with contracting
dynamics (meaning that the minimizer of the optimization problem is the equilibrium point of the dynamics):
\begin{equation*}
  \change{\min\nolimits_x}\; \mcE(x) 
  \quad\change{\longsquigglyarrow{}}\quad
  \dot{x} = \msF(x),
\end{equation*}
it is possible to use contraction theory to analyze parametric and
time-varying convex optimization, or, more precisely, parametric and
time-varying contracting dynamics:
\begin{align*}
  \change{\min\nolimits_x}\; \mcE(x,\theta) 
  \quad&\change{\longsquigglyarrow{}}\quad
  \dot{x} = \msF(x,\theta)
  \\
  \change{\min\nolimits_x}\; \mcE\bigl(x,\theta(t)\bigr) 
  \quad&\change{\longsquigglyarrow{}}\quad    
  \dot{x} = \msF\bigl(x,\theta(t)\bigr)
\end{align*}
where $\theta$ is either constant of time-varying.

Recent optimization-based control efforts~\cite{MC-ED-AB:20,AH-SB-GH-FD:21}
have focused on online feedback optimization. In this approach, they key
idea is to regulate the input-output pair $(u,y)$ of a dynamical system
subject to a disturbance $w(t)$.
\begin{figure}[h]\centering
  \includegraphics[width=.9\linewidth]{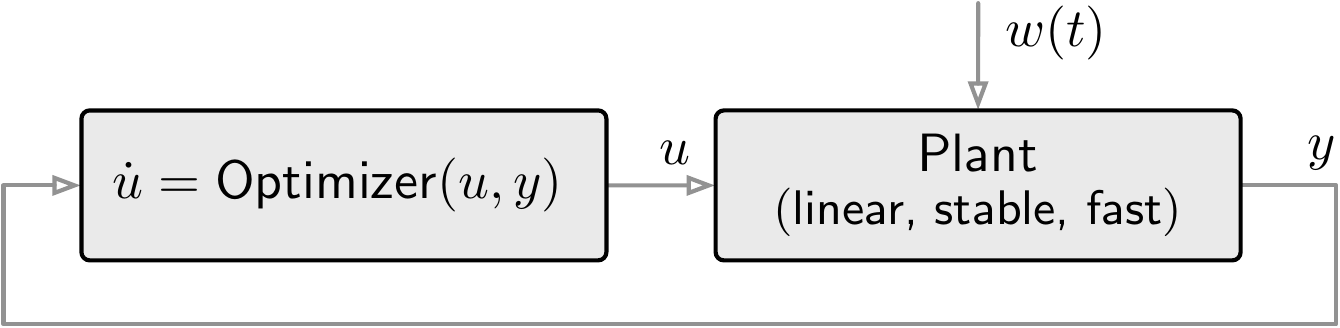}
  \caption{Online feedback optimization: typical setup and assumptions}\label{fig:feedback-optimization}
\end{figure}
\begin{figure*}[ht]\centering
	\includegraphics[width=\linewidth]{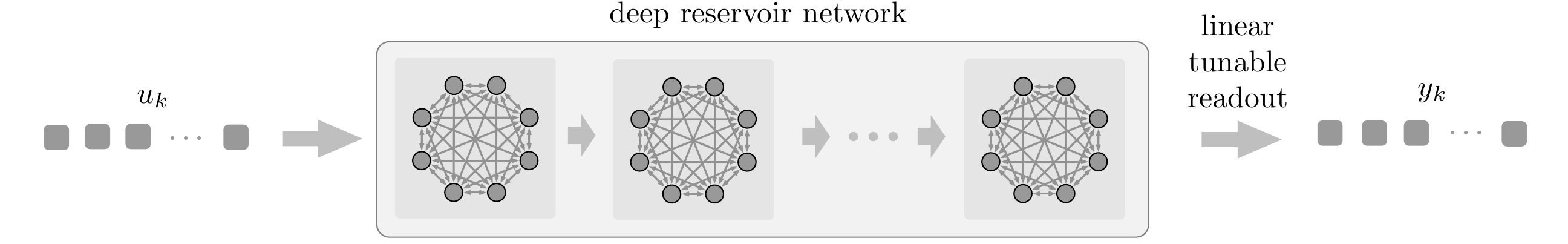}
	\caption{An example deep architecture for a reservoir
		computer~\cite{CG-AM:17}. Based upon our discussion, this deep
		reservoir network is contracting if each reservoir's synaptic matrix
		satisfies the log norm condition $\lognorm{A^{(i)}}{\infty} <
		1$.} \label{fig:nn-deepRC}
\end{figure*}

As illustrated in Figure~\ref{fig:feedback-optimization}, we aim to find
the optimal control $u$ minimizing:
\begin{equation*}
     \begin{cases}
       \change{\min\nolimits_{u,y}} \quad & \text{cost}_1(u) + \text{cost}_2(y) \\ \text{subj. to}\quad & y = \mathsf{Plant}\big(u,w(t)\big)
     \end{cases} 
\end{equation*}
Specifically, assuming a linear relationship from input $u$ and disturbance
$w$ to the output $y$, we define
  \begin{align*}
    u^*\bigl(w(t)\bigr) := \; \change{\argmin\nolimits_{u}} \;     \;&\; \phi(u) + \psi(y(t))
    \\
    \text{subj to } \;&\; y(t) = Y_u u + Y_w w(t)
  \end{align*}
  where we design $\phi$ to be strongly convex with parameter $\mu$ and
  $\psi$ to be convex.  Instead of exactly computing $u^*\bigl(w(t)\bigr)$
  at each instant of time, we define the \emph{gradient controller}
  \begin{multline*}
    \dot u = \fGradCtrl (u,w) \\ \;:= \; - \nabla_u \mcE(u,w) = - \nabla \phi(u)  - Y_u^\top \nabla \psi(Y_u u + Y_w w)
  \end{multline*}
  Since the gradient descent of a strongly convex function is strongly
  contracting (see Section~\ref{sec:Numerous dynamical systems are
    contracting}), Property~\ref{prop:iISS+eqT} implies that the gradient
  controller tracks the optimal controller satisfying:
  \begin{equation*}
    \limsup_{t \to \infty} \|u(t) - u^*\bigl(w(t)\bigr) \|
    \quad \leq \quad  \frac{\ell_w}{\nu^2} \; \limsup_{t \to \infty} \|\dot{w}(t)\|
  \end{equation*}
  where
  \begin{enumerate}
  \item $\osL_u(\fGradCtrl) \leq - \nu$, and
  \item $\Lip_w(\fGradCtrl)  = \ell_w := \norm{Y_u^\top}{} \Lip(\nabla \psi) \norm{Y_w}{}$
  \end{enumerate}
  We refer to~\cite{LC-FB-EDA:23g} for a more complete contractivity
  analysis based upon singular perturbation methods.

\smallskip\subsubsection*{Implicit and reservoir models in machine learning}
A second broad application area for contraction theory is the design of
machine learning models. The idea is to use contractivity properties to
establish accurate, reproducible, and robust behavior in face of uncertain
stimuli and dynamics.

\begin{figure}[h]\centering
  \includegraphics[width=.5\linewidth]{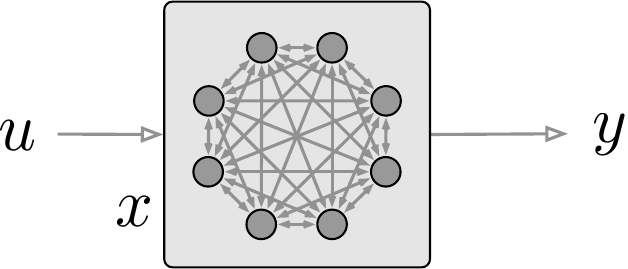}
  \caption{Implicit NNs embed feedback loops in their input-output
    relationship, unlike feedforward machine learning models. In these
    models function evaluation is performed by solving a fixed point
    equation.}\label{fig:nn-implicit}
\end{figure}

We here briefly review two models. First, \emph{implicit neural networks}
(NNs)~\cite{LEG-FG-BT-AA-AYT:21,SJ-AD-AVP-FB:21f}, also called deep
equilibrium networks~\cite{SB-JZK-VK:19}, are a class of implicit-depth
learning models that generalize feedforward neural network models and have
demonstrated improved accuracy and reduction in memory requirements, see
Figure~\ref{fig:nn-implicit}.

As starting point we consider the fixed point equation
\begin{align}
  x &= \Phi(Ax+Bu+b) \label{eq:INNs-fixedpoint}
\end{align}
whose solution is the equilibrium of both a continuous-time firing rate
model and its Euler discretization (also called leaky integrator neurons):
\begin{align}
  \dot{x} &= -x + \Phi(Ax+Bu +b) \label{eq:INNs-recurrent} \\  
  x_{k+1} &= (1-\alpha)x_k + \alpha \Phi(Ax_k + Bu+b) \label{eq:INNs-discrete}
\end{align}
In light of the previous sections and~\cite{SJ-AD-AVP-FB:21f}, if the
synaptic matrix is designed to satisfy $\mu_{\infty}(A)<1$, then (adopting the shorthand $(v)_+ := \relu(v), (v)_- := -\relu(-v)$)
\begin{itemize}
\item the implicit NN~\eqref{eq:INNs-fixedpoint} has a unique solution
  $x^*(u)$,
\item the continuous-time model~\eqref{eq:INNs-recurrent} is
  infinitesimally contracting with rate $1-(\mu_{\infty}(A))_+$,
\item the discrete-time model~\eqref{eq:INNs-discrete} is contracting with
  factor $\ds 1 - \frac{1 - (\mu_{\infty}(A))_+}{1 - \min_{i}(a_{ii})_-}$ at
  the optimal step size $\ds \alpha^*=\frac{1}{1 - \min_{i}(a_{ii})_-}$.
\end{itemize}
Additionally, the model~\eqref{eq:INNs-fixedpoint} is robust in the sense
that
\begin{equation*}
  \Lip_{u \to x} =\frac{\|B\|_{\infty}}{1 - (\mu_{\infty}(A))_+},
  \;\text{and}\; \frac{\|\Delta x^*\|_{\infty}}{\|x^*\|_{\infty}}
  \leq \frac{\|\Delta A\|_{\infty}}{1 - (\mu_{\infty}(A))_+}.
\end{equation*}

\change{Note that, starting from the fixed point
  problem~\eqref{eq:INNs-fixedpoint}, we have leveraged continuous-time
  contraction properties to establish the sharpest known conditions for
  discrete-time contraction, convergence, and robustness.}

As second machine learning application we briefly mention \emph{reservoir
computing}, illustrated in Figure~\ref{fig:nn-deepRC}.  This class of
machine learning models was introduced in~\cite{HJ:01,ML-HJ:09} to process
temporal and sequential data.  In these models, a critical property is that
the reservoir state asymptotically depends only on the input stimulus,
while the influence of initial conditions should asymptotically
vanish. This property is called the \emph{echo state property} (namely, the
state should be an echo of the input, not of the initial conditions) or the
fading memory property --- and it is an immediate consequence of
contractivity. Obtaining sharp contractivity estimates is crucial since
reservoir computers appear to work best when at the edge of stability.


\begin{figure*}[ht]\centering
	\includegraphics[width=.23\linewidth]{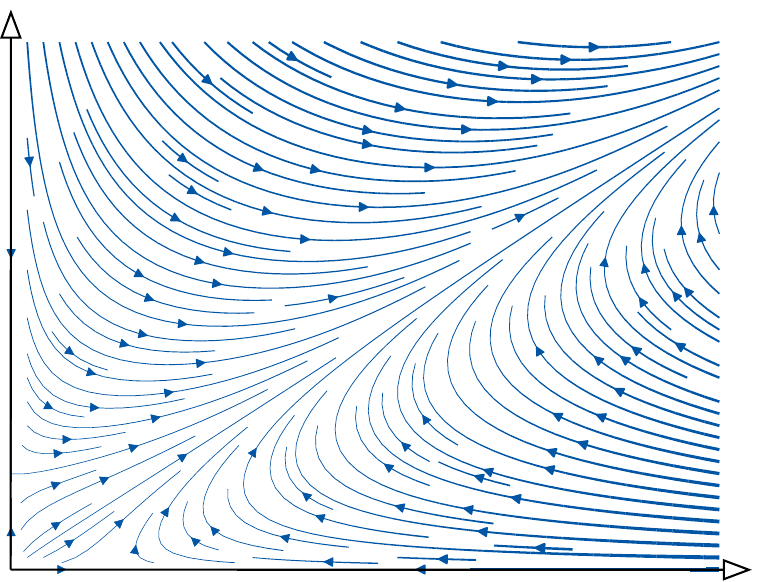}
	\hfil$\quad\vline\quad$\hfil
	\quad
	\includegraphics[width=.18\linewidth]{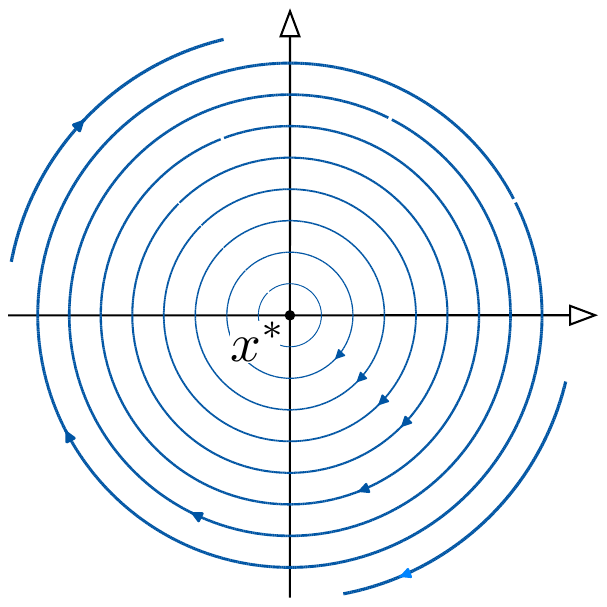}
	\includegraphics[width=.18\linewidth]{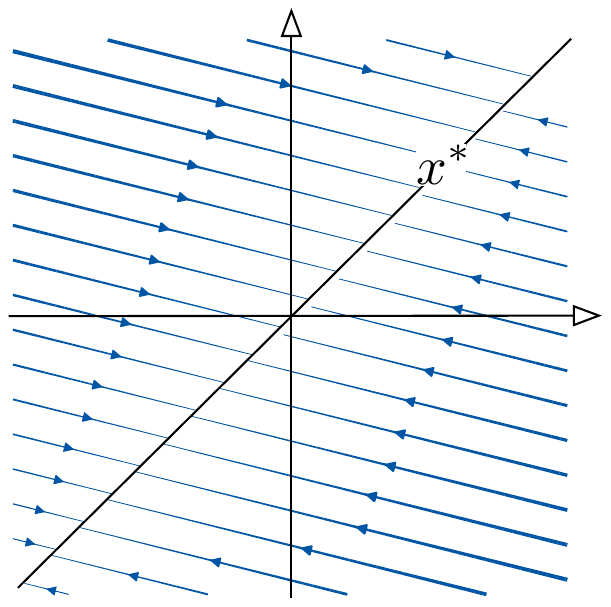}
	\includegraphics[width=.24\linewidth]{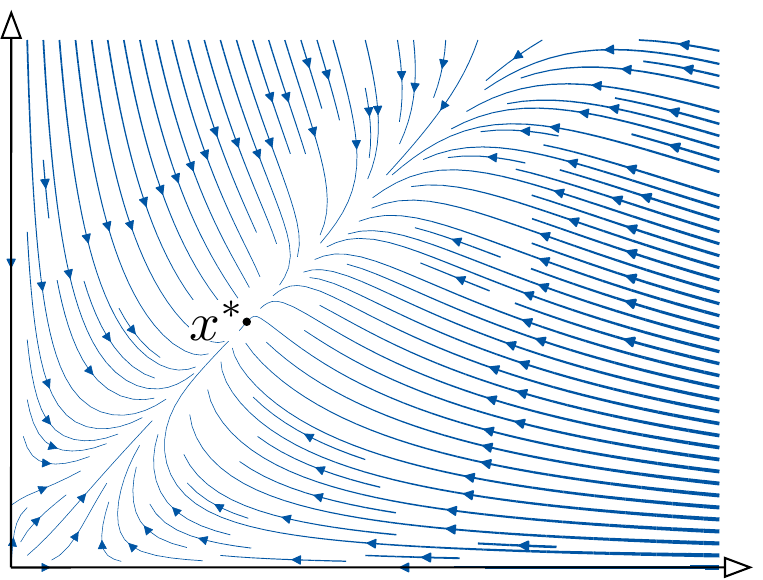}
	\caption{The dichotomy property for weakly-contracting systems: either
		each trajectory is unbounded or each trajectory is bounded. Left panel:
		every trajectory is unbounded and the system has no equilibrium. Right
		panel (with three images): every trajectory is bounded, the system has
		at either one or infinitely many equilibria and local asymptotic
		stability of an equilibrium implies global asymptotic
		stability.}\label{fig:weak-contractivity}
\end{figure*}
\smallskip\subsubsection*{Biologically-plausible competitive neural networks}
Finally, we briefly present an application to biologically-plausible neural
circuits.  Understanding the functionality of recurrent neural networks is
a major neuroscientific objective; we refer to~\cite{CP-DBC:19} for a
review of neuroscience-inspired learning and signal processing algorithms.

An important problem in this context is the study of neural networks
capable of solving dimensionality-reduction problems and, specifically,
sparse reconstruction problems.
\begin{figure}[h]\centering
	\includegraphics[width=.8\linewidth]{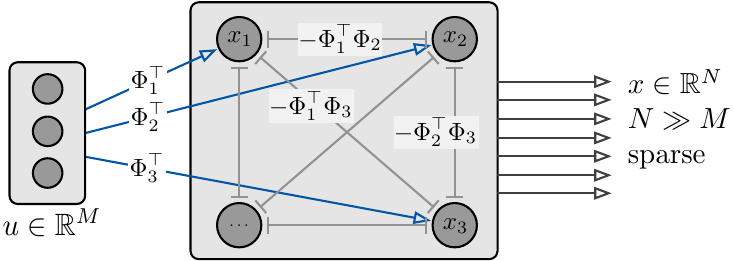}
	\caption{Diagram illustrating the positive competitive neural
		network~\eqref{eq:positive-competitive-NN}.  The competition between
		each two neurons $i$ and $j$ is proportional to $\Phi_i \cdot \Phi_j$,
		the similarity between columns $i$ and $j$.}\label{fig:nn-competitive}
\end{figure}
These problems involve approximating a given input stimulus with a set of
sparse active neurons.  Here is a simplified problem setup
from~\cite{CJR-DHJ-RGB-BAO:08,VC-AG-AD-GR-FB:23a}. Given an input stimulus
$u\in\real^M$, a vector of neural activations $x\in\real^N$, and a scalar
$\lambda$, the neural circuit aims to solve
\begin{equation}\label{eq:sparse-reconstruction-cost}
  \min_{x\in\real^N, x \geq 0} \quad  \underbrace{\tfrac{1}{2} \norm{u - \Phi x}_{2}^2}_{\text{reconstruction error}} \quad+\quad
   \underbrace{\lambda\;\norm{x}_{1}}_{\text{sparsity-promoting regularizer}}
\end{equation}
Here $\Phi\in\real^{M\times{N}}$ is a highly redundant ($N\gg M$)
nonnegative dictionary matrix with unit-norm columns ($\norm{\Phi_i}{}=1$).
Using concepts from proximal operators and contraction theory one can
design and confirm the functionality of the positive firing-rate
competitive network (see Figure~\ref{fig:nn-competitive}):
\begin{equation}
  \dot{x}_i(t) = -x_i +
  \relu\bigl(-\sum\nolimits_{j=1,j\neq{i}}^n \Phi_i^\top\Phi_j  x_j + \Phi_i^\top u -\lambda\bigr).
  \label{eq:positive-competitive-NN}
\end{equation}

Specifically, equilibrium points of~\eqref{eq:positive-competitive-NN} are minimizers of~\eqref{eq:sparse-reconstruction-cost} and if $x_i(0) \geq 0$, then $x_i(t) \geq 0$ for all $t \geq 0$. 
Moreover, under a restricted isometry property of the dictionary matrix, the dynamics are locally contracting~\cite{VC-AG-AD-GR-FB:23a}.



\myclearpage
\newcommand{\mcC}{\mathcal{C}}
\newcommand{\mcX}{\mathcal{X}}
\section{Contraction theory can be vastly extended in numerous directions}\label{sec:extensions}

While contracting dynamical systems enjoy numerous desirable properties, there are many systems that are not contracting yet enjoy similar properties. To this end, many variations on contraction theory have been proposed and in this section, we highlight three extensions to the original theory.

\subsection{Weakly Contracting Dynamics}

Systems with conserved quantities or translational invariance cannot
possibly be contracting since the system trajectories cannot fully forget
initial conditions. For example, continuous-time flow networks and
continuous-time averaging are examples of linear systems which obey this
conservation and invariance property, respectively. However, both of these
dynamics (and many others) enjoy a different property, known as weak
infinitesimal contraction (or infinitesimal nonexpansiveness), where the
distance between trajectories is nonincreasing. Stated more concretely, a
dynamical system $\dot{x} = \msF(x)$, is weakly infinitesimally contracting
if $\mu(\jac{\msF}(x)) \leq 0$ for all $x$. In analogy to standard
nonexpansive maps, weakly infinitesimally contracting dynamics induce flows
which are nonexpansive for all $t \geq 0$.

In line with the theory of nonexpansive maps, weakly contracting dynamical
systems obey a dichotomy property. Either every trajectory is unbounded
(i.e., no equilibrium point exists) or every trajectory is bounded and
there exists at least one equilibrium point~\cite{SJ-PCV-FB:19q}. Moreover,
if there exists an equilibrium point which is locally asymptotically
stable, then it is also globally asymptotically stable. See
Figure~\ref{fig:weak-contractivity} for examples of weakly contracting
dynamical systems.

\subsection{Locally Contracting Dynamics}

Although it is desirable to ensure global contraction, in many practical
examples, dynamics are only contracting in a region (possibly, but not
necessarily, containing an equilibrium point). For example, gradient
descent for nonconvex objective functions is locally contracting in a
neighborhood of a local minimizer if and only if the objective function is
strongly convex in the same neighborhood.

\begin{figure}[ht]\centering
  \includegraphics[width=.8\linewidth]{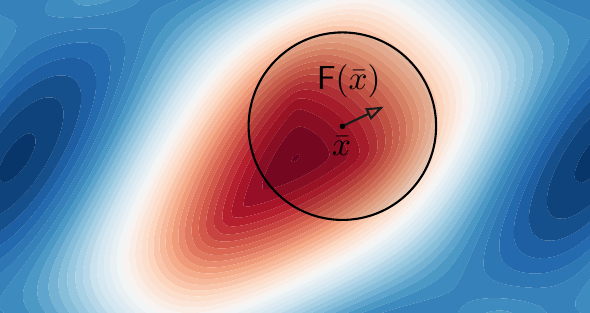}
  \caption{Illustration of a locally contracting vector field.  Given a
    vector field $\msF$ on a planar domain and a norm, the contour plot
    shows values of the quantity $\mu(\jac{\msF})$: red values correspond
    to regions where the vector field $\msF$ is strongly infinitesimally
    contracting $\mu(\jac{\msF}(x))<0$ and blue values correspond to
    regions where $\mu(\jac{\msF}(x))>0$. To establish the existence of a
    forward-invariant set, it is useful to find a closed ball centered at
    $\overline{x}$ inside the red region, with small
    residual~$\norm{\msF(\overline{x})}{}$.} \label{fig:local-contractivity}
\end{figure}

In short, if there exists a convex forward-invariant set, $\mcC$, for the
dynamical system, $\dot{x} = \msF(x)$, then all the standard consequences
of contractivity hold for all trajectories remaining inside the set, e.g.,
the existence of a unique equilibrium point $x^* \in \mcC$. The challenge
is thus in finding a set, $\mcX$, for which $\mu(\jac{\msF}(x)) \leq -c$
for all $x \in \mcX$ and then finding a forward invariant set, $\mcC
\subseteq \mcX$. In Figure~\ref{fig:local-contractivity}, we plot an
instance of a locally contracting system.

\subsection{Contraction Theory on Manifolds}

\begin{figure}[h]\centering
  \includegraphics[width=.8\linewidth]{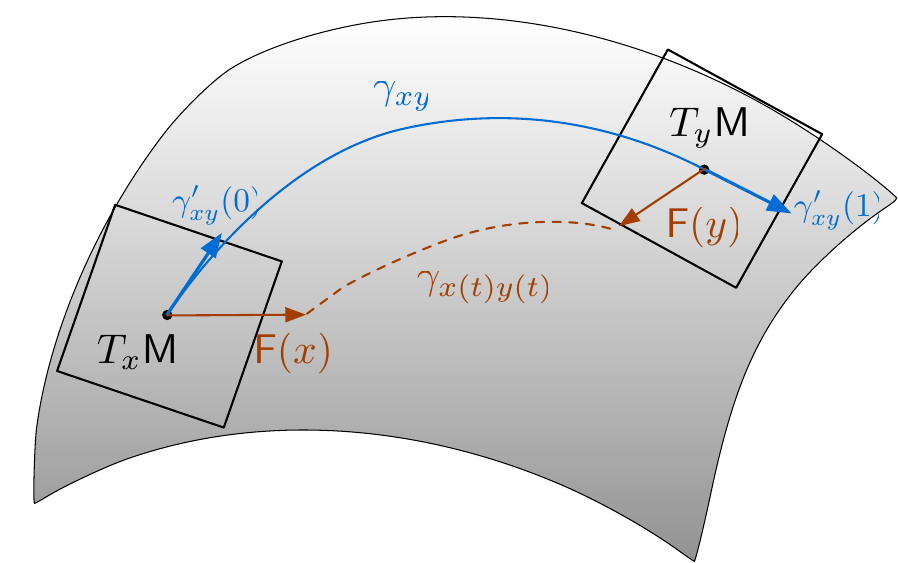}
  \caption{Contractivity of a vector field $\msF$ on a Riemannian manifold:
    the length of the geodesic curve $\gamma_{xy}$ connecting any two
    points $x$ and $y$ decreases along the flow of $\msF$, as a function of
    the inner product between $\msF$ and the geodesic velocity vector at
    $x$ and $y$.} \label{fig:contraction-Riemannian}
\end{figure}

Contraction theory on Riemannian manifolds was initiated by the seminal
work~\cite{WL-JJES:98}. In brief summary, given a system $\dot{x} =
\msF(x)$ with $x \in \real^n$, it is proposed to give $\real^n$ the
structure of a Riemannian manifold by introducing a mapping
$\map{M}{\real^n}{\real^{n \times n}}$ and two positive constants $a_0,a_1
> 0$ satisfying $M(x) = M(x)^\top$ and $a_0 I_n \preceq M(x) \preceq a_1
I_n$ for all $x \in \real^n$. Then, $M(x)$ induces a state-dependent norm
on $\real^n$, $\|\cdot\|_{M(x)}$ or, more precisely, a state-dependent
inner product $\inprod{\cdot}{\cdot}_{M}$ and Riemannian metric. The
dynamical system is contracting with respect to this Riemannian metric with
rate $c > 0$ if
\begin{equation}\label{eq:Riemannian}
  M(x) \jac{\msF}(x) + \jac{\msF}(x)^\top M(x) + \dot{M}(x) \preceq -2cM(x),
\end{equation}
for all $x \in \real^n$. Here, the quantity $\dot{M}$ is a shorthand for
the Lie derivative of $M$ along trajectories of $\dot{x} = \msF(x)$.  The
state-dependent LMI~\eqref{eq:Riemannian} is a generalization of the
classical result by Demidovich. In the context of log norms, if $M(x) =
\Theta(x)^\top \Theta(x)$, the LMI~\eqref{eq:Riemannian} is equivalent to
asking that
\begin{equation*}
  \mu_2\Bigl(\Theta(x)\jac{\msF}(x)\Theta(x)^{-1} + \dot{\Theta}(x)\Theta(x)^{-1}\Bigr) \leq -c.
\end{equation*}
In short, contraction with respect to a Riemannian metric ensures that the
geodesic distance between any two trajectories is exponentially
decaying. See Figure~\ref{fig:contraction-Riemannian} for an
illustrative example of contraction on a Riemannian manifold.

Since the seminal work by Lohmiller and Slotine, much interest has focused
on contraction theory on manifolds. In~\cite{JWSP-FB:12za}, a formal
coordinate-free analysis was provided. In~\cite{FF-RS:14}, the theory was
extended to Finsler manifolds, \change{which serves as a non-Euclidean analog to contraction with respect to Riemannian metrics}. Finally, in~\cite{IRM-JJES:17}, the theory
was extended to design controllers such that the closed-loop system is
guaranteed to be contracting with respect to a certain Riemannian metric.

\myclearpage
\section{Conjectures and Future Directions}\label{sec:open-problems}
We conclude this opinion paper with some open problems in theory,
applications, and computation.

\smallskip\subsubsection*{Theory}
While the theoretical underpinnings of contraction theory with respect to
norms and Riemannian metrics are largely understood, there remain many
fundamental open problems surrounding sharpness of contraction conditions
and their extensions. For example, sharp characterizations of contractivity
exist for some special dynamical systems (e.g., gradient flow, firing rate
neural networks, and certain Lur'e models), yet there are other relatively
simple dynamical systems whose sharpest rates of contraction are still
unknown, e.g., primal dual dynamics for linear equality-constrained
minimization and Hopfield neural networks with diagonally stable synaptic
matrices.

At the same time, a comprehensive understanding of the recent theoretical
extensions of $p$-dominance~\cite{FF-RS:19,YS-YK-NW:23}, $k$- and
$\alpha$-contraction~\cite{CW-RP-MM-JJES:20,CW-IK-MM:22,EBS-OD-MM:23} is
still missing along with rich sets of examples and necessary and sufficient
conditions for establishing these properties.

Additionally, an input-output perspective on contracting systems appears
largely missing. Namely, comparisons between contracting systems and
incrementally passive~\change{\cite{AP-LM:08}} or dissipative systems~\change{\cite{FF-RS:19}} could yield novel insights.
In a similar vein, since contracting systems have properties akin to those
of stable linear systems (apart from the superposition principle), it is
natural to ask whether contracting systems admit frequency-domain analysis
tools (see the early work~\cite{AP-NVDW-HN:07} for ideas).

A final theoretical topic of interest is developing novel mathematical
tools to establish contractivity. For contraction with respect to Euclidean
norms, a useful tool is the S-Lemma~\cite{IP-TT:07} which provides
conditions under which one quadratic inequality (e.g., contractivity) is
implied by other quadratic inequalities (e.g., slope constraints on
nonlinearities). For contraction with respect to non-Euclidean norms, a
non-Euclidean S-Lemma was recently proposed in~\cite{AVP-AD-FB:22a} and
necessary and sufficient conditions for zero duality gap were shown in the
case of the $\ell_1$ norm with Metzler matrices.  It remains an open
problem whether such results can be extended to other important cases such
as contractivity with respect to the $\ell_\infty$ norm. Developments in
this direction may rely upon the theory of integral linear
constraints~\cite{CB:13}.

\smallskip\subsubsection*{Applications}
In Section~\ref{sec:modern-applications}, we presented three modern
applications where contraction theory plays a pivotal role, namely
optimization-based control, machine learning, and dynamical neuroscience.
In each of these applications, there remains room for improvement. For
example, contraction-theoretic tools are only recently coming to bear on
the topics of suboptimal model predictive control and in control barrier
functions. Specifically, the incremental input-to-state stability and
equilibrium tracking properties may be used to provide novel robustness
bounds to inaccurate solvers or uncertainty in the dynamics model.

Beyond these three applications, we list in the Introduction several other
research areas where contractivity tools play an important role.  For
example, in the study of biochemical reaction networks, convergence and
robustness properties are being established independently of specific forms
of kinetics by leveraging contraction with respect to polyhedral
norms~\cite{FB-GG:14,MAAR-DA-EDS:23,AD-EDS:24}. Additionally, in the study
of traffic networks, weak contraction with respect to the $\ell_1$ norm is
adopted to establish the existence of steady state free-flow
equilibria~\cite{GC-EL-KS:15,SC-MA:15}.

In short, we believe that contraction theoretic ideas can be leveraged to
analyze and design dynamical systems and algorithms in many disparate
application domains.

\smallskip\subsubsection*{Computation}

In establishing contraction of a given dynamical system or closed-loop
control system, the typical challenge is in finding with respect to what
norm or Riemannian metric the system is contracting. Due to recent
improvements in machine learning and numerical optimization solvers,
several software packages have become available to either (i) numerically
estimate contraction metrics, (ii) design controllers so that the
closed-loop system is approximately contracting, or (iii) establish
robustness guarantees of neural networks via a contraction analysis. For
example, \cite{PZ-AL-KA-AG-MP-NH:22,HT-SJC:21} feature various versions of
SOS-programming, gridding and interpolation, and neural network-based
software for the computation of control contraction
metrics. \cite{NHB-MR-RW-JC-IRM:23} provides a Julia package for neural
networks and data-driven control with robustness guarantees ensured by
contraction. Finally, \cite{SJ-AD-AVP-FB:21f} contains software to train
implicit neural networks which are guaranteed to be contracting with
respect to an $\ell_\infty$ norm.

Despite this progress, no single software library is comprehensive and
widely accepted as the reference implementation in this area. An open
challenge and burgeoning area of software engineering would be to create a
robust codebase for establishing contractivity of structured nonlinear
systems leveraging recent advances in nonlinear programming. An important
extension to this codebase would be the ability to design controllers with
guarantees of closed-loop contractivity.


\smallskip\subsubsection*{Closing Thoughts}
In this opinion paper, we have extolled the virtues of contraction theory
and highlighted its universality and applicability. Specifically, we have
argued that (i) contraction theory is a unifying framework emerging from
classical and modern works, (ii) contractivity is a
computationally-friendly, robust, and modular stability, (iii) numerous
dynamical systems are contracting, (iv) contraction theory is relevant to
modern applications, and (v) contraction theory can be vastly extended in
numerous directions.

As a takeaway message for control theoreticians and engineers, we hope we
have convinced them to:
\begin{quote}
(1) search for contraction properties, \\
(2) design engineering systems to be contracting, and \\
(3) verify correct and safe behavior via known \\ \phantom{(4)} Lipschitz and contractivity constants.
\end{quote}

\myclearpage
\section{Acknowledgments}
We would like to thank our coauthors and colleagues: Veronica Centorrino,
Pedro Cisneros-Velarde, Sam Coogan, Lily Cothren, Emiliano Dall'Anese, Giulia
De~Pasquale, Robin Delabays, Xiaoming Duan, Anand Gokhale, Saber Jafarpour, Michael Margaliot, Ron Ofir,
Anton Proskurnikov, Giovanni Russo, Francesco Seccamonte, John Simpson-Porco, Kevin Smith, and
Maria Elena Valcher for all of their stimulating conversations.

\begin{arxiv}
  \bibliographystyle{plainurl+isbn}
\end{arxiv}
\begin{lcss}
  \bibliographystyle{plainurl+isbn}
\end{lcss}
\bibliography{alias, Main, FB, New}

\end{document}